\documentclass[a4paper,UKenglish,cleveref, autoref, thm-restate]{lipics-v2021}
\usepackage{enumitem}
\usepackage{tabularx}
\usepackage{array}
\usepackage[table]{xcolor}
\usepackage{makecell}
\usepackage{fontawesome5}

\usepackage{framed}
\definecolor{mycolor}{gray}{0.9}

\newcommand{\infobox}[1]{%
\noindent
\begingroup
\setlength{\fboxsep}{0.5em}%
\setlength{\fboxrule}{0.7pt}%
\fcolorbox{black}{mycolor}{%
\parbox{\dimexpr\linewidth-2\fboxsep-2\fboxrule}{%
\vspace{0.1ex}
#1
\vspace{0.1ex}
}%
}%
\endgroup
}

\newcolumntype{L}[1]{>{\raggedright\arraybackslash}p{#1}}
\newcolumntype{C}[1]{>{\centering\arraybackslash}p{#1}}

\title{\fontsize{16}{18}\selectfont
Emerging Challenges in Threat Modeling for GenAI-Augmented Systems: A View from the Trenches}

\titlerunning{Emerging Challenges in Threat Modeling for GenAI-Augmented Systems} 

\author{Nicolás E. {Díaz Ferreyra}}{Institute of Software Security, Hamburg University of Technology, Germany \and \url{http://www.tuhh.de/softsec} }{nicolas.diaz-ferreyra@tuhh.de}{https://orcid.org/0000-0001-6304-771X}{}

\author{Manish Mahesh Kumar}{Institute of Software Security, Hamburg University of Technology, Germany \and \url{http://www.tuhh.de/softsec} }{manish.manish.mahesh.kumar@tuhh.de}{}{}

\author{Nohemí Villarreal}{CREATUM GmbH, Germany \and \url{http://www.creatum.io/} }{nohemi.villarreal@creatum.io}{}{}

\author{Pankaj Pantel}{CREATUM GmbH, Germany \and \url{http://www.creatum.io/} }{pankaj.patel@creatum.io}{}{}

\author{Immo Brueggemann}{CREATUM GmbH, Germany \and \url{http://www.creatum.io/} }{immo.brueggemannl@creatum.io}{}{}

\author{Riccardo Scandariato}{Institute of Software Security, Hamburg University of Technology, Germany \and \url{http://www.tuhh.de/softsec} }{riccardo.scandariato@tuhh.de}{https://orcid.org/0000-0003-3591-7671}{}

\authorrunning{N.E. Díaz Ferreyra et al.} 

\Copyright{Jane Open Access and Joan R. Public} 



\EventEditors{}
\EventNoEds{2}
\EventLongTitle{2026 International Symposium on Empirical Software Engineering and Measurement}
\EventShortTitle{ESEM 2026}
\EventAcronym{ESEM}
\EventYear{2026}
\EventDate{October 4--9, 2026}
\EventLocation{Munich, Germany}
\EventLogo{}
\SeriesVolume{}
\ArticleNo{}

\begin{document}
\nolinenumbers

\maketitle

\begin{abstract}

Threat modeling remains a central task in secure software engineering, as it enables the identification of security issues from system architectures. As Generative Artificial Intelligence (GenAI) becomes increasingly pervasive across software systems, traditional threat modeling methods (e.g., STRIDE) are insufficient to assess emerging GenAI-specific risks. In this work, we present the first results from an exploratory assessment of GenAI-aware threat modeling methods in a Small and Medium Enterprise (SME) setting. For this, we conducted a rapid literature review to select relevant techniques and systematically applied three shortlisted methods to an industrial case study involving a GenAI-augmented system. The results highlight differences in the threats identified by each technique and reveal limited support for certain GenAI-specific risk categories, particularly those related to software supply chains and human-centered security issues. We further report practitioners' perceptions of the usability and integration of these methods in SME development workflows, including their perceived effort and adoption challenges.

\end{abstract}

\section{Introduction}
\label{sec:intro}

The security of software-intensive systems is a multifaceted challenge that requires coordinated efforts from stakeholders across the entire development lifecycle. Threat Modeling (TM) is a key component of the security toolkit, as it provides a structured set of methods and artifacts to support the early identification of potential security threats at the design stage~\cite{van2022descriptive}. Over the years, several architecture-centered TM methods have been proposed and extensively discussed in the current literature \cite{xiong2019threat}. One of the most widely adopted is STRIDE \cite{howard2006security}, a methodology introduced by Microsoft in 2008, which enables practitioners to systematically identify threats by categorizing them into six classes: (S)poofing, (T)ampering, (R)epudiation, (I)nformation Disclosure, (D)enial of Service, and (E)levation of Privilege \cite{naik2024comparative}. Similar approaches also rely on mnemonic threat taxonomies and structured representations of system components (e.g., data flow diagrams) to reason about potential attack vectors and security violations \cite{sion2020security,mbaka2025assessing,lohman2023review}. Furthermore, dedicated tool support is often available to facilitate their application, thereby reducing manual effort \cite{granata2024systematic}.

\textbf{\textsl{Motivation.}} General-purpose methods such as STRIDE often struggle to capture the full threat landscape of complex, domain-specific systems. Consequently, several variants have emerged over the years to address the specific security challenges of domains such as IoT \cite{srikumar2022striped}, automotive systems \cite{abuabed2023stride}, and industrial control systems \cite{sassnick2024stride} among others. These variants typically extend existing taxonomies with additional threat categories or introduce domain-specific modeling constructs to better represent the system under analysis. The rapid adoption of Generative Artificial Intelligence (GenAI) technologies has further motivated the adaptation of State-of-the-Art (SotA) TM techniques to account for risks specific to these systems, such as prompt injection and data leakage through model interactions \cite{mauri2021stride,grosse2024towards,von2024asset,jedrzejewski2025thremolia}. Still, there is limited empirical evidence on the applicability of these techniques in real-world settings. In particular, it remains unclear how well these approaches support practitioners, especially in Small and Medium Enterprise (SME) contexts, and where their most salient gaps and challenges lie \cite{jedrzejewski2024threat}.

\textbf{\textsl{Contribution and Research Questions.}} In this work, we explore the gaps and emerging challenges of TM techniques when applied to GenAI-augmented software systems. To this end, we first conducted a rapid literature review to identify SotA methods tailored to the analysis of GenAI-specific security threats. We then applied three shortlisted techniques to an industrial case study provided by a local SME to examine their performance in practice. Finally, we complemented our analysis with a practitioner survey to capture perceptions regarding the usability and integration of these techniques in real-world development workflows. Overall, we aim to answer the following Research Questions (RQs):

\begin{itemize}[topsep=0pt]
    \item \textit{\textbf{RQ1}: How effective are state-of-the-art threat modeling techniques in identifying GenAI-specific security issues in an SME setting?} To answer this RQ we modeled and analyzed the system architecture of the given case study following the steps prescribed by each of the three selected TM methodologies. The outputs of each method (e.g., type and number of security threats) were then assessed in terms of their coverage against the OWASP Top 10 for LLM Applications \cite{owasp_llm_top10}, a taxonomy of threats specific to Large Language Models (LLM) systems developed by security practitioners.
    \item \textit{\textbf{RQ2}: How do practitioners perceive the adoption and integration of these techniques in an SME context?} We conducted a follow-up survey with members of the development team of the SME that provided the case study, aiming to elicit their perceptions of each of the applied TM techniques. The survey included a set of items assessing dimensions such as perceived usefulness (e.g., ability to identify relevant and actionable threats), effort required for regular use, and integration with current tools and practices. Thereby, we provide initial insights into practitioners' perceptions, including practical gaps, challenges, and barriers to adoption and integration.
   
\end{itemize}







\section{Background and Related Work}


\hspace{\parindent} \textbf{\textsl{(i) Threat modeling in practice.}} Threat Modeling (TM) is central to many security-by-design frameworks and standards, as it supports the systematic identification of risks arising from system architectures and information flows. Typically, it involves modeling the target system, eliciting threats using structured taxonomies, and prioritizing them to inform mitigation strategies. Data Flow Diagrams (DFDs) are frequently used to create a graphical, high-level representation of the software architecture and depict how information moves across the main system components \cite{tuma2021finding}. In their most basic variant, DFDs include four types of elements, namely (i) external entities, (ii) data flows, (iii) processes, and (iv) data stores \cite{schneider2024dataflow}. During threat analysis, security and domain experts inspect each of these elements with the help of heuristics provided by the TM method to identify potential security threats. In STRIDE, these heuristics take the form of a structured mapping between DFD elements and threat categories (e.g., data stores associated to information disclosure threats)~\cite{howard2006security}.




Because of its relative ease of use and adaptability, STRIDE has become a de facto baseline in security TM and has been evaluated across multiple studies. These evaluations have highlighted not only its lightweight design but also practical limitations regarding scalability and applicability to complex systems \cite{granata2024systematic,xiong2019threat}. In a nutshell, architecture-centered TM methods like STRIDE require significant manual effort and expert knowledge to model the system under analysis with sufficient detail and extract meaningful security threats \cite{mbaka2025assessing}. Scandariato et al. \cite{scandariato2015descriptive} assessed the performance of STRIDE through a descriptive study and concluded that, despite its lightweight approach, it is relatively time-consuming and prone to \textit{false negatives} (i.e., threats that analysts fail to identify). Recent work by Mbaka et al. \cite{mbaka2025assessing} investigated the role of DFDs and supplementary material in validating the completeness of a STRIDE-driven threat analysis. Overall, their findings suggest that case study descriptions and threat descriptions are sometimes more useful than DFDs for assessing the coverage of a given analysis output. Nevertheless, proper tool support remains a critical factor for the usability of TM techniques, especially in collaborative team settings \cite{bernsmed2022adopting}.

\textbf{\textsl{(ii) GenAI-specific security challenges.}} As GenAI becomes increasingly central to modern software architectures, it introduces new security threats that require systematic identification and mitigation \cite{wang2025unique}. The OWASP Top 10 for LLM Applications \cite{owasp_llm_top10} provides a suitable reference taxonomy for characterizing these threats, covering issues related to the use of malicious prompts, mishandling sensitive data, and adversarial model behavior. Prompt injection emerges at the top of the list, representing the use of carefully crafted inputs to override LLMs' instructions and trigger unintended or unsafe behavior. Next comes sensitive information disclosure, a critical security issue that can affect both users and the LLMs themselves, for instance, when personal information or technical details (e.g., fragments of LLMs' source code) are leaked in their output \cite{das2025security}. The use of third-party pretrained LLMs can also introduce severe upstream vulnerabilities (e.g., biased outputs or system failures) if their security is compromised. Such vulnerabilities are often the consequence of data or model poisoning attacks, in which malicious actors manipulate the training data or the model itself to influence the behavior of downstream applications \cite{zhang2024imperceptible,cotroneo2024vulnerabilities}. LLMs are also prone to producing inaccurate or flawed outputs, especially when applied to complex tasks such as source code development \cite{chen2025evaluating}. Hence, over-reliance on these systems and granting them excessive agency can introduce significant security risks, particularly when model outputs are accepted or executed without sufficient human oversight or validation \cite{diaz2026concerns}.

Recent TM approaches have begun incorporating AI- and GenAI-specific security threats into their analysis frameworks. Mauri and Damiani~\cite{mauri2021stride} extended STRIDE with AI-specific assets distributed across the ML lifecycle, namely data collection, model training, and deployment. Their method defines \textit{failure models} for each asset category and derives the corresponding threats from violations of the associated security properties (e.g., authenticity, integrity, and confidentiality). Similarly, von der Assen et al. \cite{von2024asset} introduced \textit{ThreatFinderAI}, an approach geared towards contemporary AI and LLM-driven architectures in which assets such as pre-trained models and prompts are explicitly represented and analyzed. Threat identification is then supported through the construction of \textit{attack graphs} based on prospective risk scenarios collected from multiple authoritative AI-security knowledge bases, including OWASP AI Exchange~\cite{owaspai2026} and MITRE ATLAS~\cite{mitreatlas2026}. Still, to the best of our knowledge, empirical evaluations of these and other methods alike remain limited, especially regarding their application to real-world scenarios and industrial settings. Moreover, little is known about practitioners' acceptance of these techniques, the effort required to integrate them into existing development workflows, and the practical frictions that may hinder their regular use.




\enlargethispage{2\baselineskip}
\section{Research Methodology}

\begin{figure}[t]
    \centering
\includegraphics[width=0.8\linewidth]{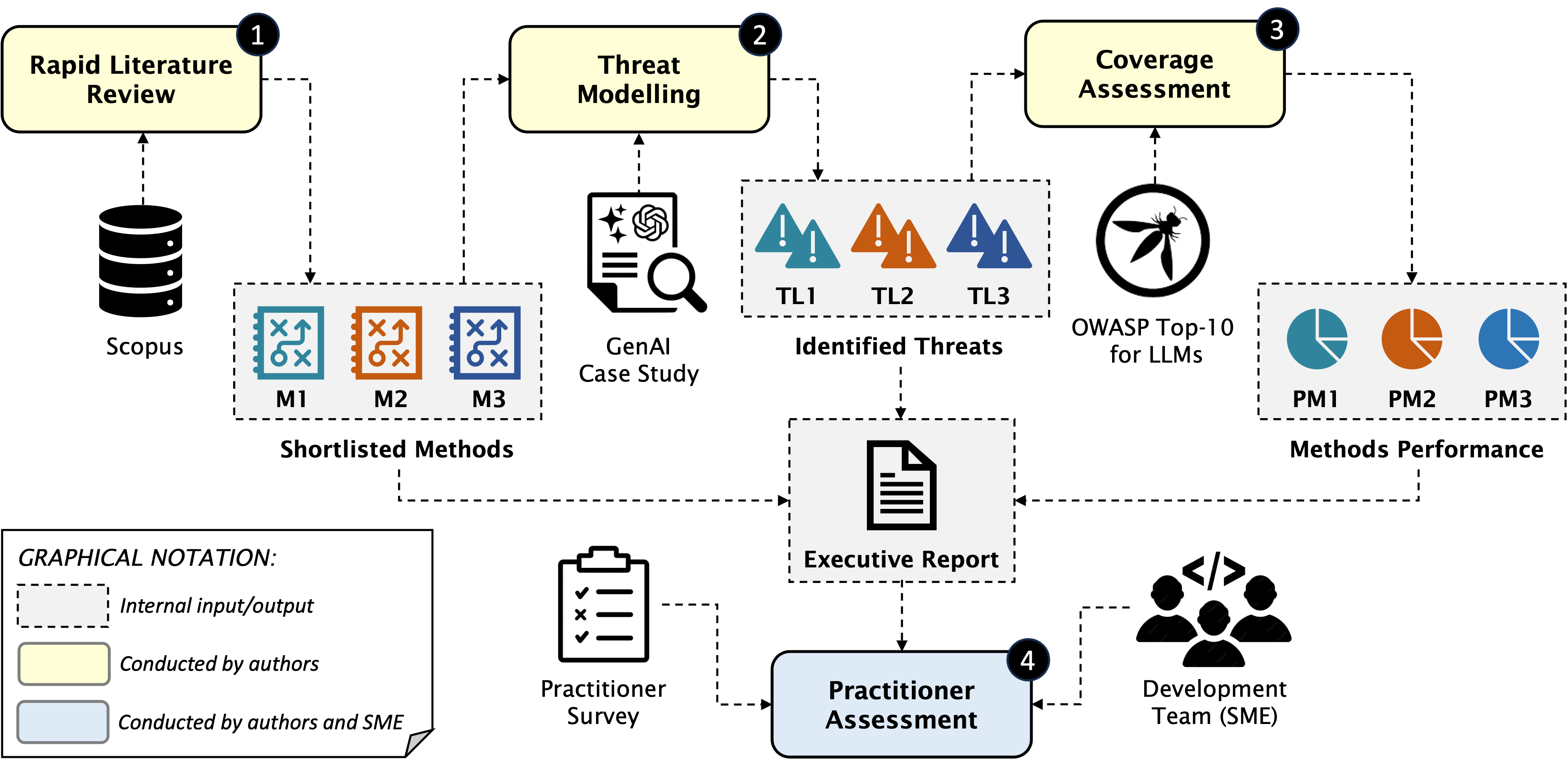}
    \caption{Applied research methodology.}
    \label{fig:method}
    \vspace{-2ex}
\end{figure}

Fig.~\ref{fig:method} illustrates the methodology we applied to address the RQs proposed in Section~\ref{sec:intro}. We first conducted a Rapid Literature Review (RLR) to identify SotA TM techniques suitable for the analysis of security threats in GenAI-augmented systems (STEP 1). We shortlisted three candidate methods based on their maturity level (e.g., validation extent) and applied them to a case study provided by a local SME (STEP 2). The performance of each method was then assessed regarding its coverage of security threats included in the OWASP Top-10 for LLMs~\cite{owasp_llm_top10} relevant to such a case study (STEP 3). It is worth noting that the SME development team was kept informed of the progress at each method step before compiling the results (i.e., the RLR, the TM analysis, and the coverage assessment) into an Executive Report (ER). Once finalized, the team was then asked to review the ER and complete a survey on their perceptions of each method's suitability, the actionability of its outputs, and the anticipated integration effort in the current SME development setting (STEP 4). In the following subsections, we provide further details on each methodological step and the supporting artifacts employed throughout the study.


\subsection{Rapid Literature Review}

Rapid Literature Reviews (RLRs) provide a lightweight framework for characterizing the SotA within a limited time frame while still following a structured and transparent process~\cite{pizard2025using}. In contrast to more exhaustive evidence synthesis methods, RLRs typically involve narrower search scopes and simplified screening procedures, making them suitable for exploratory and early-stage studies. Given the emerging nature of GenAI-oriented TM research and the exploratory goals of this work, we considered an RLR appropriate for identifying candidate methods applicable to GenAI-augmented systems.

\begin{figure}[t]
    \centering
    \includegraphics[width=0.9\linewidth]{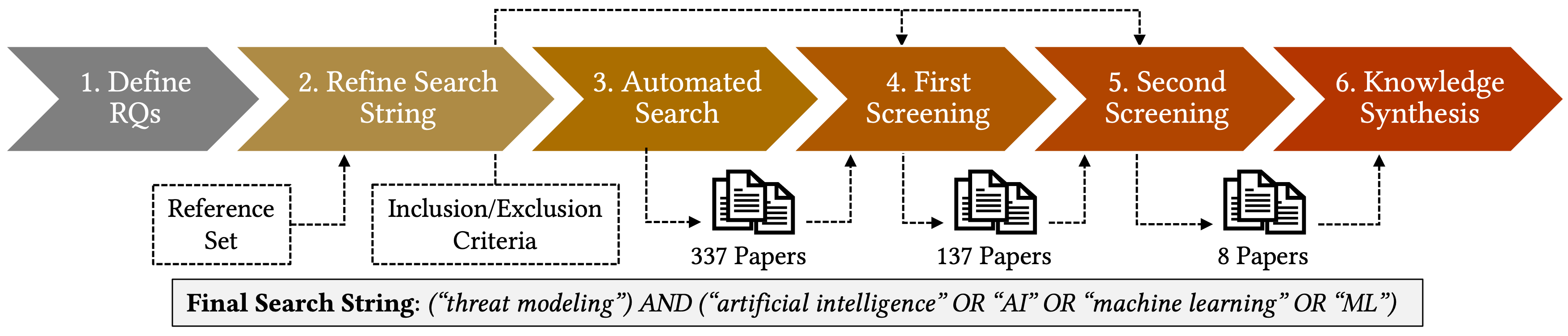}
    \caption{Applied Rapid Literature Review (SLR) methodology (STEP 1).}
    \label{fig:rapid_review}
\end{figure}

Fig.~\ref{fig:rapid_review} illustrates the steps we conducted through the RLR. We first started defining the RQs that would guide the literature search. Particularly, we centered around the following RLR-specific RQ: \textit{``What threat modeling techniques documented in the current literature are suitable for the analysis of GenAI-augmented systems?''}. Next, we defined the search engine and keyword string employed for the automated literature search. We selected Scopus\footnote{http://www.scopus.com} as the primary search engine due to its broad coverage of major software engineering publication venues, including ACM Digital Library and IEEE Xplore. To refine the search string, we constructed a small reference set of representative publications on TM for GenAI-augmented systems (i.e., \cite{mauri2021stride,von2024asset}) and iteratively adjusted the search terms until all reference studies were retrieved. The final query is depicted at the bottom of~Fig.~\ref{fig:rapid_review}. 

The automated search was conducted in June 2025 and returned 337 candidate studies, which were subsequently processed during the screening stages. We retained studies that (i) described architecture-centered TM techniques, (ii) considered security threats relevant to GenAI-based systems, (iii) were written in English, and (iv) were published at well-established software engineering or security venues. These inclusion and exclusion criteria were first applied to the title and abstract of the candidate studies during the initial screening stage, resulting in 137 potentially relevant papers. The same criteria were then applied to the full text of these studies during a second screening stage, ultimately leading to the selection of 8 primary studies that underwent a knowledge synthesis process. For this, we
followed a lightweight open coding approach in which one author extracted emerging themes and patterns relevant to the proposed RLR-specific RQ. Another author reviewed these themes afterward and, if discrepancies arose, these were resolved through a negotiated agreement.

\subsection{Methods Selection}

The thematic analysis of the selected primary studies surfaced a set of recurring dimensions characterizing the identified TM methods. First, we analyzed the \textit{System Type} targeted by each method, namely whether the technique was designed for traditional ML-based systems or more contemporary GenAI-oriented architectures involving LLMs and transformer-based components. Second, we examined the prescribed \textit{Modeling Approach}, including the architectural representation and modeling artifacts employed throughout the TM process (e.g., traditional or custom DFD syntax). Third, we assessed the extent of \textit{Tool Support}, particularly whether the method provided (semi-) automated support for threat identification. Finally, we considered the reported \textit{Evaluation} strategy of each method, including proof-of-concept applications, case studies, and developer-centered usability evaluations. 

\begin{table*}[ht]
\caption{Characterization of the selected primary studies.}
\centering
\small

\resizebox{0.95\linewidth}{!}{%
\begin{tabularx}{\linewidth}{
|>{\scriptsize\arraybackslash}X
|>{\scriptsize\centering\arraybackslash}p{0.04\linewidth}
|>{\scriptsize\centering\arraybackslash}p{0.14\linewidth}
|>{\scriptsize\centering\arraybackslash}p{0.20\linewidth}
|>{\scriptsize\centering\arraybackslash}p{0.05\linewidth}
|>{\scriptsize\centering\arraybackslash}p{0.15\linewidth}|}
\hline

\rowcolor{black}
\textcolor{white}{\textbf{Source}} &
\textcolor{white}{\textbf{Year}} &
\textcolor{white}{\textbf{System Type}} &
\textcolor{white}{\textbf{Modeling Approach}} &
\textcolor{white}{\textbf{Tool}} &
\textcolor{white}{\textbf{Evaluation}}
\\
\hline

\rowcolor{white}
Von der Assen et al. \cite{von2024threatfinderai} & 2024  & GenAI & Asset-Enriched DFD & YES & Case study \\
\hline

\rowcolor{gray!25}
Von der Assen et al. \cite{von2024asset} & 2024 & GenAI & Asset-Enriched DFD & YES & Case study \\
\hline

\rowcolor{white}
Mauri and Damiani \cite{mauri2021stride} & 2021 & ML & Asset-Enriched DFD & NO  & Case study \\
\hline

\rowcolor{white}
Mauri and Damiani \cite{mauri2022modeling} & 2022 & ML & Asset-Enriched DFD & NO & Case study \\
\hline

\rowcolor{white}
Wilhjelm and Younis \cite{wilhjelm2020threat} & 2020 & ML & DFD & NO & Case study \\
\hline

\rowcolor{gray!25}
G\"{u}len et al. \cite{gulen2024threat} & 2024 & GenAI & DFD & NO & Proof of concept \\
\hline

\rowcolor{white}
Messas et al. \cite{messas2024saife} & 2024 & ML & Unspecified & NO & Usability study \\
\hline

\rowcolor{gray!25}
Kumar et al. \cite{kumar2024} & 2024 & GenAI & DFD & NO & Case study \\
\hline

\end{tabularx}
}
\label{tab:classification}
\end{table*}
Table~\ref{tab:classification} summarizes the characterization of the eight primary studies. The rows highlighted correspond to the TM methods selected for their application to the case study. 
Our selection criteria centered on techniques (i) geared towards GenAI-enriched system architectures and (ii) validated through case studies. Since these criteria narrowed the selection to only two methods (i.e., \cite{von2024asset,gulen2024threat}), we decided to relax the latter and include one additional TM approach (i.e., \cite{kumar2024}). This resulted in the following TM techniques being selected for application to the case study:
\begin{itemize} [topsep=0pt]
    \item \textbf{M1} - \textit{AI-as-a-Service (AIaaS) Framework} \cite{gulen2024threat}. Relies on DFD-based system representations and structures the analysis around asset groups such as \textit{Data}, \textit{Model}, \textit{Environment}, and \textit{Process}. For each asset category, the framework provides STRIDE-based threat listings along with prospective mitigation actions.
    \item \textbf{M2} - \textit{Attacks on Dataset, Model and Input (ADMIn) Framework} \cite{kumar2024}. It adopts an attack-centric perspective around three major attack surfaces, namely \textit{Datasets}, \textit{Models}, and \textit{System inputs}. The analysis is structured through DFD-like architectural representations and predefined attack categories associated with each attack surface. The method further provides guidance on threat prioritization and mitigation activities.
    \item \textbf{M3} - \textit{ThreatFinderAI} \cite{von2024asset}. This approach extends traditional DFDs with AI-specific assets (e.g., prompts and pre-trained models), which are subsequently analyzed through attack graphs and risk scenarios. Threat identification is semi-automated through the integration of external security knowledge bases such as MITRE ATLAS \cite{mitreatlas2026} and OWASP AI Exchange \cite{owaspai2026}. Unlike M1 and M2, this method is tool-supported.
\end{itemize}

\subsection{SME-Provided Case Study}

\begin{figure}[t]
    \centering
    \includegraphics[width=0.75\linewidth]{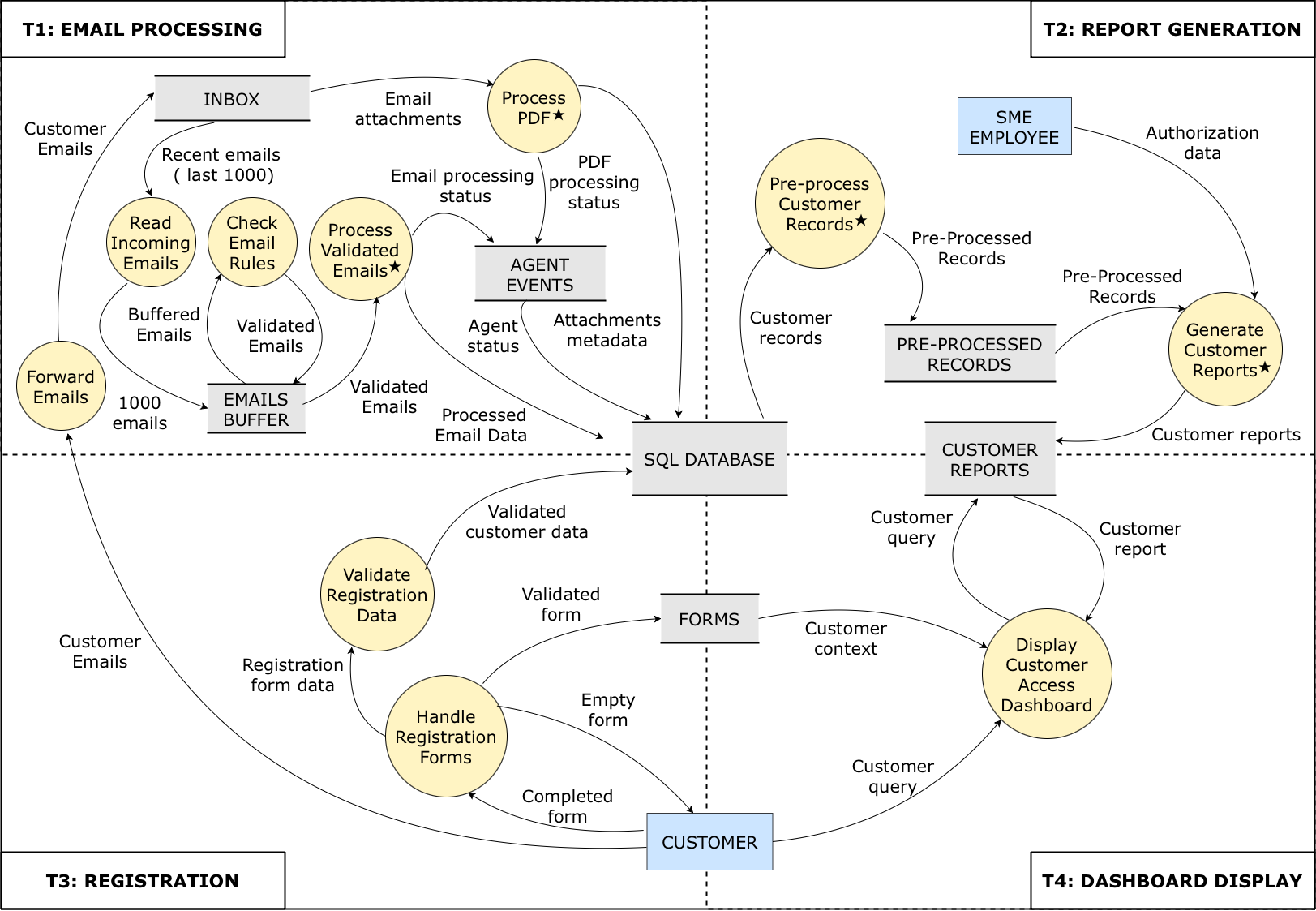}
    \caption{Data Flow Diagram (DFD) of the analyzed case study.}
    \label{fig:case_study}
\end{figure}

Fig.~\ref{fig:case_study} illustrates a DFD of the SME-provided case study. It describes the main architectural components (i.e., data stores, processes, data flows, and entities) of a business intelligence system that automatically processes e-mails from customers and generates customized market reports. It encompasses four major tasks, namely e-mail processing (T1), report generation (T2), customer registration (T3), and dashboard display (T4). Such tasks (T1 and T2 in particular) are partially supported by LLM agent processes (marked with a $\star$) in charge of (i) processing incoming emails and their attachments (i.e., PDF files), (ii) validating such emails (e.g., checking for duplicates), (iii) pre-processing internal records (e.g., customer profiles and historic data), and (iv) generating reports tailored to the requirements of each client. From a technological perspective, the system leverages Microsoft Power Platform\footnote{https://www.microsoft.com/en-us/power-platform} services for workflow orchestration and customer data management, while Azure OpenAI\footnote{https://azure.microsoft.com/en-us/products/ai-services/openai-service} is employed to support the LLM-based processing components embedded within the architecture. The generated business intelligence reports and dashboards are further integrated with Power BI services for visualization and customer-facing reporting activities. Overall, the analyzed architecture comprises 10 processes, including 3 LLM-assisted components, as well as 7 data stores, 27 data flows, and 2 external entities.





\subsection{Practitioner Survey}

Seven members of the SME development team participated in a survey to identify potential adoption and integration challenges associated with the evaluated TM techniques. Particularly, they were first asked to review the content of the ER, namely a description of the three shortlisted TM methods, their output for the given case study, and their coverage regarding the OWASP Top-10 for LLMs. Then, we asked them to provide their perspectives on each method by assessing a set of statements across seven dimensions, including the method \textit{effectiveness} (i.e., \textit{``The method helped identify important security threats for our system''}), \textit{actionability} (i.e., \textit{``The method provided clear mitigation actions for addressing the identified threats''}), and \textit{awareness} (i.e., \textit{``The method surfaced threats we would likely have missed otherwise''}), among others. Each participant assessed each statement three times (i.e., for each method) using 6-point Likert values (i.e., \textit{completely disagree}; \textit{disagree}; \textit{somewhat disagree}; \textit{somewhat agree}; \textit{agree}; \textit{completely agree}). The survey concluded with the following open-ended questions aimed at collecting further insights into the added value, limitations, and adoption challenges of GenAI-aware TM techniques:
\begin{itemize}[topsep=0pt]
    \item \textbf{OE1}: \textit{What were the top 3 most valuable threats identified across the methods, and why?}
    \item \textbf{OE2}: \textit{What would have to change for this method to be adoptable in your workflow?}
\end{itemize}

\vspace{2ex}
\infobox{\noindent\faFolderOpen ~All study materials are available in the paper's \textbf{Replication Package} (Section~\ref{sec:data}), including the RLR protocol, the corresponding code book, the survey instrument, the aggregated survey results, and the OWASP Top-10 for LLMs mapping.}



\section{Results}

In the following subsections, we report the findings obtained from the assessment of the selected TM methodologies (Section~\ref{sec:comparison}) and the practitioner survey (Section~\ref{sec:survey}). We narrowed the former to the GenAI-aided processes within the system, along with their corresponding incoming and outgoing data flows. Thereby, we focus the analysis on the system components directly influenced by GenAI capabilities and, therefore, more likely to expose GenAI-specific security threats. In particular, we considered the following processes and data flows:

\begin{itemize}[topsep=0pt]
    \item \textbf{\textsl{P1: PDF Processing}}, with incoming flow \textit{IF-1.1: Email attachments} and outgoing flows \textit{OF-1.1: PDF processing status} and \textit{OF-1.2: Attachment metadata}.
    
    \item \textbf{\textsl{P2: Validated Email Processing}}, with incoming flow \textit{IF-2.1: Validated emails} and outgoing flows \textit{OF-2.1: Processed email data} and \textit{OF-2.2: Email processing status}.
    
    \item \textbf{\textsl{P3: Customer Report Generation}}, with incoming flows \textit{IF-3.1: Authorization data} and \textit{IF-3.2: Pre-processed records} and outgoing flow \textit{OF-3.1: Customer reports}.
    
    \item \textbf{\textsl{P4: Customer Record Pre-Processing}}, with incoming flow \textit{IF-4.1: Customer records} and outgoing flow \textit{OF-4.1: Pre-processed records}.
\end{itemize}

\subsection{RQ1: Effectiveness of Threat Modeling Techniques}


Although all three TM techniques addressed some GenAI-related security threats, they differed substantially in terms of abstraction level, threat granularity, and analysis focus. M1~\cite{gulen2024threat} adopts a STRIDE-inspired approach, while incorporating security threats traditionally associated with adversarial ML \cite{lin2021adversarial}, including \textit{model poisoning}, \textit{model inversion}, \textit{evasion}, \textit{membership inference}, and \textit{model extraction} attacks. In this context, \textit{model inversion} and \textit{membership inference} refer to attempts to infer sensitive information (e.g., data used for fine-tuning) through interactions with the LLM-based components, whereas \textit{evasion attacks} aim to manipulate externally controlled inputs to alter model behavior or bypass intended processing logic. While \textit{data poisoning} (i.e., injecting malicious or manipulated data into LLM-based workflows) is not explicitly represented in the taxonomy, it can still be operationalized through classical STRIDE categories such as tampering or manipulation during data transmission. As shown in Table~\ref{tab:owasp_coverage}, these threats, which emerge from all GenAI-aided processes and incoming data flows, largely align with two of the OWASP Top-10 threat categories for LLMs, namely \textit{Sensitive Information Disclosure} and \textit{Data and Model Poisoning}.


Compared to M1, M2~\cite{kumar2024} more explicitly captures GenAI-specific runtime threats, particularly by incorporating \textit{prompt injection} in addition to traditional adversarial ML attacks. These threats, which directly correspond to OWASP's \textit{Prompt Injection} category, were considered relevant to all analyzed GenAI-aided processes. Furthermore, the method identifies outgoing data flows as potential targets of membership inference attacks (referred to as ``data exfiltration'' attacks), which, similarly to M1, are conceptually related to OWASP's \textit{Sensitive Information Disclosure} category. Certain outgoing data flows, particularly operational status flows such as \textit{OF-1.1} and \textit{OF-2.2}, could not be naturally mapped to this category as they do not expose, in principle, semantically rich and potentially sensitive information. 


M3 \cite{von2024asset} was the only approach providing dedicated tool support, together with a custom syntax and semantics for representing AI-related architectural components and threats. Consequently, the analysis was conducted on a different but semantically-equivalent DFD, requiring the identified threats to be mapped back to the original processes and data flows. Although this mapping was not always one-to-one, it remained relatively straightforward in practice (e.g., \textit{data ingestion} were generalized to all incoming data flows), leading to results largely comparable to those of M2 (Table~\ref{tab:owasp_coverage}). Despite their differing abstractions and threat modeling strategies, all three methods identified processes \textit{P1} and \textit{P3} as potential targets of \textit{Denial-of-Service (DoS)} attacks, mainly due to their reliance on computationally intensive LLM operations and externally controlled inputs. These threats are conceptually related to OWASP's \textit{Unbounded Consumption} category, as attackers may attempt to trigger excessive inference operations or resource exhaustion scenarios through malicious or oversized inputs. Nevertheless, the evaluated methods showed limited support for several OWASP Top-10 for LLMs categories, particularly those associated with agentic behavior, system prompt exposure, and unsafe downstream integration of generated outputs.

\vspace{1ex}
\infobox{\noindent\faSearch~\textbf{Findings RQ1.} Although the evaluated TM techniques identified several GenAI-related issues, their coverage was strongest for input- and model-interaction threats (e.g., \textit{Prompt Injection}, \textit{Data and Model Poisoning}). Coverage was limited or absent for categories requiring broader architectural, supply-chain, or human-centered reasoning (e.g., \textit{Supply Chain Vulnerabilities} and \textit{Excessive Agency}). M3 provided the broadest support due to its tool-supported analysis and dedicated AI-oriented modeling syntax.
}

\newcommand{\owaspitem}[2]{\hspace{0.1em}\makebox[1em][r]{#1.}\ #2}

\begin{table*}[t]
\scriptsize
\centering
\caption{Coverage of OWASP Top-10 for LLMs across the evaluated TM methods}
\label{tab:owasp_coverage}
\renewcommand{\arraystretch}{1.2}
\setlength{\tabcolsep}{2.5pt}

\resizebox{0.95\linewidth}{!}{%
\begin{tabularx}{\linewidth}{|l|>{\centering\arraybackslash}X|>{\centering\arraybackslash}X|>{\centering\arraybackslash}X|}
\hline\rowcolor{black}
\textcolor{white}{\textbf{OWASP Top-10 for LLMs}} & \textcolor{white}{\textbf{M1}} & \textcolor{white}{\textbf{M2}} & \textcolor{white}{\textbf{M3}} \\
\hline

\owaspitem{1}{Prompt Injection}
& -
& P1,P2,P3,P4
& P1,P2,P3,P4 \\ \hline

\owaspitem{2}{Sensitive Information Disclosure}
& P1,P2,P3,P4
& OF-1.2,OF-2.1,OF-3.1
& P1,P2,P3,P4,\mbox{OF-1.2}, \mbox{OF-2.1},\mbox{OF-3.1} \\ \hline

\owaspitem{3}{Supply Chain Vulnerabilities}
& -
& -
& - \\ \hline

\owaspitem{4}{Data and Model Poisoning}
& P1,P2,P3,P4,IF-1.1, \mbox{IF-2.1},IF-3.1,IF-3.2, \mbox{IF-4.1}
& P1,P2,P3,P4,IF-1.1, \mbox{IF-2.1},IF-3.1,IF-3.2, \mbox{IF-4.1}
& IF-1.1,IF-2.1,IF-3.1, \mbox{IF-3.2},IF-4.1, \mbox{OF-1.2},OF-2.1,OF-3.1 \\ \hline

\owaspitem{5}{Improper Output Handling}
& -
& -
& - \\ \hline

\owaspitem{6}{Excessive Agency}
& -
& -
& - \\ \hline

\owaspitem{7}{System Prompt Leakage}
& -
& -
& - \\ \hline

\owaspitem{8}{Vector and Embedding Weaknesses}
& N/A
& N/A
& N/A \\ \hline

\owaspitem{9}{Misinformation}
& -
& -
& - \\ \hline

\owaspitem{10}{Unbounded Consumption}
& P1,P3
& P1,P3
& P1,P3 \\ \hline
\end{tabularx}%
}
\end{table*}

\subsection{RQ2: Practitioner Perceptions}\label{sec:comparison}

\begin{figure}
    \centering
    \includegraphics[width=0.9\linewidth]{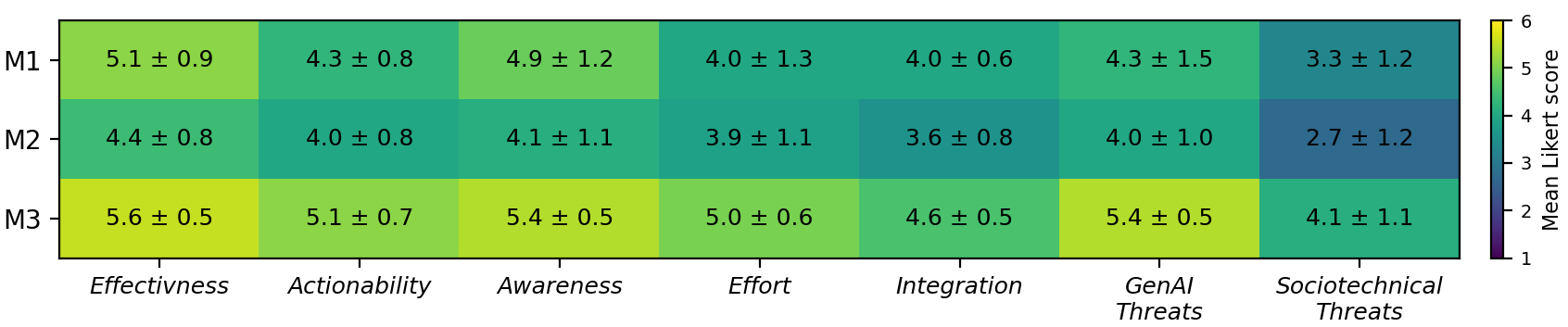}
    \caption{Practitioner ratings across M1 (\textit{AIaaS}), M2 (\textit{ADMIn}), and M3 (\textit{ThreatFinderAI}).}
    \label{fig:heatmap}
\end{figure}

Fig.~\ref{fig:heatmap} shows participants' average scores and standard deviations for each of the seven survey dimensions and TM methods. Overall, M3 received the highest scores across all dimensions, with particularly strong results regarding its capability to identify GenAI-specific security threats (\textit{GenAI Threats}) and the effort required for regular use (\textit{Effort}). This may be attributed to M3's tool support and its incorporation of external, GenAI-specific security knowledge in the underlying threat catalog. Conversely, M1 and M2 received comparable scores regarding effort, which may be explained by their lack of automation.

Both M1 and M3 received relatively higher scores than M2 in terms of identifying system-relevant threats (\textit{Effectiveness}) and security issues that participants would otherwise have overlooked (\textit{Awareness}). This may be associated with the limited relevance of one of M2's main attack surfaces (i.e., ``Models'') in this particular case, since the SME does not develop its own ML models but rather relies on externally provided LLM services. Furthermore, both M1 and M3 provide a more fine-grained catalog enriched with detailed threat descriptions, affected assets, and, in the case of M3, also prescriptive appraisal mechanisms. This may also have affected practitioners' perception of M2's support for deriving clear mitigation actions (\textit{Actionability}) and integrating the resulting outputs into existing development workflows (\textit{Integration}), which were rated lower than those of M1 and M3.

In line with the mapping results (Section~\ref{sec:comparison}), all methods received their lowest scores in identifying socio-technical security threats. This is consistent with Table~\ref{tab:owasp_coverage}, where none of the evaluated methods surfaced threats related to OWASP's \textit{Excessive Agency} category, which is the most closely related to socio-technical concerns in our analysis. Still, the slightly higher scores of M1 and M3 may be explained by their explicit consideration of potential threat actors, which can help practitioners reason about who may exploit a given threat.

When analyzing participants' answers to OE1, we observed that they consistently identified \textit{prompt injection}, \textit{data poisoning}, and \textit{sensitive information disclosure} as the most valuable threats surfaced by the methods. These were perceived as particularly relevant because the system processes external, potentially untrusted inputs, while also handling sensitive customer information: \textit{``Prompt injection was a real eye-opener. Data poisoning too---silent and sneaky. And data leakage through cloud APIs, honestly we wouldn't have caught that on our own.''} (P6). On the other hand, responses to OE2 emphasized that adoption would require lightweight integration into existing development routines, including ticketing workflows and clear ownership. Furthermore, they underscored the need for structured templates (e.g., checklists) and short training sessions on system modeling and GenAI security threats: \textit{``Templates would make it so much easier. A quick onboarding session for the team. And just connect it to our tickets---that alone would make a huge difference.''} (P5).

\vspace{1ex}
\infobox{\noindent\faSearch~\textbf{Findings RQ2.} Practitioners rated M3 highest across all survey dimensions, particularly for its support in surfacing \textit{GenAI Threats} and its relatively low perceived \textit{Effort}. M1 and M3 received higher scores than M2 for \textit{Effectiveness}, \textit{Awareness}, \textit{Actionability}, and \textit{Integration}. Across methods, \textit{Socio-technical Threats} received the lowest scores, consistent with the RQ1 findings. Participants also emphasized the need for better integration with existing development workflows, especially issue-tracking and security review processes.
}

\subsection{Threats to Validity} 

The findings of this study should be interpreted in light of its exploratory nature. \textit{External validity:} The assessment was conducted on a single case study, which limits the generalizability of the results to other GenAI-augmented systems, domains, or organizational settings. \textit{Construct validity:} The evaluated TM techniques differ in their abstraction levels, modeling assumptions, and expected inputs, making direct comparison challenging. To mitigate this issue, we normalized the analysis around GenAI-aided processes, their associated data flows, and the OWASP Top-10 for LLMs as a common reference baseline. \textit{Internal validity:} The mapping of method-specific threats to OWASP categories may have been influenced by the researchers' interpretation. Therefore, we followed a conservative mapping strategy and avoided assigning categories when the correspondence was not sufficiently clear. \textit{Conclusion validity:} the practitioner survey involved a small number of participants from the involved development team, who assessed the methods based on the Executive Report rather than by applying the techniques themselves. Hence, the survey results should be understood as initial perceptions of usefulness, actionability, and adoption challenges rather than as evidence of actual long-term adoption in practice.

\section{Prospective Research Pathways} \label{sec:survey}

In principle, the study results suggest that current TM techniques are useful for identifying GenAI-related threats, such as \textit{prompt injection}, \textit{data poisoning}, and \textit{sensitive information disclosure}. However, they show limited support for some critical threat categories, particularly those directly associated with (i) the software supply chain and (ii) the interaction between LLMs and developers. As third-party GenAI solutions become an integral part of system architectures, TM methods should provide means for identifying security risks stemming from LLM dependencies \cite{williams2025research}. Still, none of the three TM techniques we assessed offered a threat catalog and modeling syntax enriched with explicit supply-chain, GenAI-specific elements (e.g., third-party ML libraries, container images) that could help practitioners reason about such risks. Furthermore, in line with prior work \cite{bernsmed2022adopting}, the survey results indicate that tool support should not be neglected, as M3 received the highest overall perception scores among participants. This motivates the following Research Pathway (RP):

\vspace{1ex}
\infobox{\noindent\faMapSigns~\textbf{RP1}: GenAI-aware TM approaches should explicitly incorporate knowledge of supply-chain security into their frameworks while supporting some degree of automation.} 
\\

Prior work on the automatic extraction of security-enriched DFDs from source code~\cite{schneider2023automatic} could, in principle, provide a promising basis for this direction. In particular, these extraction capabilities could be extended to cover supply-chain artifacts such as LLM-relevant dependency manifests, SBOMs, and container definitions. Alongside this, recent work by Jedrzejewski et al.~\cite{jedrzejewski2025thremolia} introduced a prompt-driven threat analysis concept that leverages LLMs with RAG to generate and maintain threat models over time. As vibe coding begins to displace traditional programming practices, such approaches could help align TM with emerging AI-assisted development workflows. This integration could be further supported by structured prompt templates and checklists, as suggested by participants in the survey:

\vspace{1ex}
\infobox{\noindent\faMapSigns~\textbf{RP2}: Future work should explore prompt-guided and checklist-supported workflows for integrating GenAI-aware TM into LLM-assisted development practices.}
\\

Finally, the limited coverage of human-centered security threats across all assessed TM techniques is concerning. Prior work has shown that LLMs may produce unreliable or hallucinated results under certain conditions; therefore, GenAI-aided processes require careful scrutiny before their outputs are integrated into downstream development activities. At the same time, multiple studies raise concerns about developers' over-reliance on such outputs and reduced critical engagement during their operationalization \cite{diaz2026concerns}. This calls for TM approaches that explicitly capture (i) where flawed LLM-generated content may propagate across GenAI-augmented architectures, (ii) which actors are responsible for validating or acting upon them, and (iii) situations in which over-reliance may become security-threatening. Furthermore, suitable mitigation actions should go beyond traditional offline training and combine online behavioral interventions \cite{brown2019digital} with interactive, personalized security assistance \cite{tony2022conversational} to support practitioners when engaging with LLM-generated outputs.

\vspace{1ex}
\infobox{\noindent\faMapSigns~\textbf{RP3}: GenAI-aware TM approaches should account for developers' over-reliance on LLM-generated outputs by modeling human-centered threat propagation paths and identifying intervention points across the system architecture.}

\section{Conclusion}

TM remains a cornerstone of secure software engineering, enabling practitioners to systematically identify and mitigate security risks at the architectural level. However, as GenAI becomes increasingly embedded into modern software systems, traditional TM techniques require new abstractions and supporting mechanisms to account for the distinctive characteristics of LLM-based applications. In this paper, we presented emerging results from an exploratory assessment of three GenAI-aware TM techniques in an industrial SME case study. Our findings indicate that, although current techniques can capture prominent GenAI-related threats such as prompt injection and data poisoning, they provide only limited support for broader architectural, supply-chain, and human-centered risks.

Beyond threat coverage, our results highlight the importance of practical adoption. The survey results indicate that practitioners value TM techniques that are easy to integrate into existing development workflows while requiring minimal additional effort. As software development continues to evolve toward increasingly AI-assisted and prompt-driven practices, next-generation TM techniques should not only broaden their coverage of GenAI-specific risks but also facilitate their seamless incorporation into modern software development pipelines. Altogether, these findings suggest that the future of GenAI-aware TM depends on combining comprehensive threat coverage with lightweight, tool-supported, and actionable integration into software development practice.










\section{Data Availability Statement}\label{sec:data}

All study materials are available in the following link to the paper's \textbf{Replication Package}: \url{https://doi.org/10.5281/zenodo.21706295}. The package includes the RLR and extraction forms, the characterization of the selected TM methodologies, the survey instrument, aggregated survey results, and the OWASP Top-10 for LLMs mapping produced during the analysis.

\section{Acknowledgments}

This work was supported by the European Union under grant No. 101120393 (Sec4AI4Sec).



\bibliography{references}

\end{document}